\documentclass[
  aps,
  prd,
  reprint,
  superscriptaddress,
  amsmath,
  amssymb,
  nofootinbib,
  longbibliography
]{revtex4-2}

\usepackage{graphicx}
\usepackage{xcolor}
\usepackage{ulem}

\usepackage{mathrsfs}
\usepackage{comment}
\usepackage{hyperref}
\hypersetup{
  colorlinks=true,
  linkcolor=blue,
  citecolor=blue,
  urlcolor=blue
}

\begin{document}

\title{Near-Extremal Black-Hole Formation Observed from  Null Infinity}

\author{Thomas M\"adler}
\affiliation{Facultad de Ingenier\'ia, Universidad San Sebasti\'an,
Bellavista 7, Recoleta, Chile}
\email{thomas.maedler@uss.cl}

\author{Radouane Gannouji}
\affiliation{Instituto de F\'isica,
Pontificia Universidad Cat\'olica de Valpara\'iso,
Avenida Brasil 2950, Valpara\'iso, Chile}
\email{radouane.gannouji@pucv.cl}

\author{Steven L. Liebling}
\affiliation{Long Island University,
720 Northern Boulevard, Brookville, New York 11548, USA}

\author{Emanuel Gallo}
\affiliation{Universidad Nacional de Córdoba, Facultad de Matemática, Astronomía, Física y Computación, Grupo de Relatividad y Gravitación; Córdoba, Argentina.}
\affiliation{Instituto de F\'isica Enrique Gaviola,
CONICET, C\'ordoba, Argentina}

\date{\today}

\begin{abstract}
We identify near-extremal black-hole formation directly from observables at future null infinity. Fully nonlinear evolutions of the spherically symmetric Einstein–Maxwell–charged scalar system in \(3+1\) dimensions form black holes with charge-to-mass ratios up to \(0.993\) from regular, horizon-free, superextremal initial data. Quasinormal frequencies and redshift decay independently recover the near-extremal charge-to-mass ratio. Critical mass scaling is recovered at null infinity and charge scaling at the apparent horizon. The charged scalar Newman–Penrose quantity obeys an asymptotic balance law that replaces the conservation law of the neutral theory.
\end{abstract}

\maketitle

\textit{Introduction.---}
The four laws of black hole mechanics arose as a remarkable analogy with the laws of thermodynamics~\cite{Bardeen:1973gs}.
This analogy became solid when Hawking discovered that the surface gravity, $\kappa_H$, of a stationary black hole relates to the black body radiation emission temperature of  quantum fields near a black hole horizon~\cite{Hawking:1975vcx}.
While  laws zero, one, and two
 have been proven rigorously by numerous authors, the validity of the third law was debated \cite{Farrugia_1979,Israel:1986gqz,Dadhich:1997rq,Racz:2000af} and thereby re-formulated in many ways in the last 50 years. 
One of its formulations states that a so-called extremal black hole with $\kappa_H=0$ cannot be reached by a finite sequence of operations and thus requires an infinite time to form.
The classical proof of the third law \cite{Israel:1986gqz} assumes a connected apparent horizon in its construction  and is based on the idea that any overcharging or overspinning of a black hole would violate the weak energy condition. 
But, recently, by allowing the apparent horizon to be disconnected, Kehle and Unger~(KU)~\cite{Kehle:2022uvc} have demonstrated rigorously that the third law is false. 
It is  a peculiarity of disconnectedness of the apparent horizon that  `pushes' the black hole to be extremal with $\kappa_H=0$ and thereby to lose all trapped surfaces near the horizon. 
The simplest realization of an extremal black hole is  the  spherically symmetric extremal Reissner Nordstr\"om~(RN) black hole for which its mass  equals its charge.

Their seminal work~\cite{Kehle:2022uvc}   employed the novel method of characteristic gluing, which consists of gluing spacetime invariants of different spacetimes along a null hypersurface~\cite{2024AnHP...25.3081A,2023CMaPh.403..275A}.   
KU's disproof is  teleological in nature and it  constructs the null hypersurface of an emerging  event horizon for a dynamically  forming extremal RN black hole  with three different pieces. 
The null hypersurface has a regular vertex in Minkowski space, an intermediate piece containing the (matter) data responsible for forming the black hole, and an asymptotic part which is isometric to an extremal RN black hole. 

An important part of KU is the choice of initial data of the intermediate part.  
The work of~\cite{Gadioux:2025unn} provided more detail on the form of initial data required, and they numerically investigated the gluing along an RN horizon.
The initial data  consists of disjoint pulses which serve to charge the black hole to extremality. 
Earlier simulations of charged-scalar field collapse using a \(3+1\)  formulation found subextremal RN black hole from superextremal initial data \cite{Torres:2014fga}.
Other numerical studies of the third law
include:   investigations in 4+1 de Sitter and vacuum spacetimes \cite{Dunajski:2026qip,Crump:2026kgu}, charged scalar field collapse in $4+1$ dimensional spacetime~\cite{Mittal:2026ryg}, and the evolution of charged particles~\cite{East:2025nfb}.

 A violation of the third law implies that it should be possible to observe (near) extremal black hole formation from regular initial data in finite (computational) time at future null infinity, $\mathfrak I^+$. 
Up to now, (near-)extremal black-hole formation has been identified only from horizon quantities in the bulk \cite{Dunajski:2026qip,Crump:2026kgu,Gelles:2026bix,Mittal:2026ryg,East:2025nfb}, not from radiation at $\mathfrak I^+$.

Here we report on the results of a numerical study of the collapse of a charged scalar field in
{\it four dimensional}, spherically symmetric spacetime covering both the vertex of  the null cones and future null infinity. For sufficiently large initial data,
a black hole forms with properties   
determined from global observables, such as redshift, news function, Bondi mass, and Bondi charge. 
Most importantly, we find regions of parameter space in which a nearly extremal black hole
forms.
Our results also include the extraction of scaling laws near the threshold for black hole formation as well as the determination of quasinormal modes of near-extremal black holes at future null infinity.
We work with geometric  Gaussian units $G=c=1$ with $\kappa = 8\pi$ as the gravitational constant.

\textit{Model and formulation.---}
We consider Einstein gravity coupled to a massless complex scalar field
$\Phi$ coupled to a Maxwell field $A_a$ with gauge charge $q$.  The action is
\begin{equation}
S=\int d^4x\,\sqrt{-g}
\left[
\frac{R}{2\kappa}
-\frac{F_{ab}F^{ab}}{16\pi}
-\frac{1}{2}\overline{D_a\Phi}D^a\Phi
\right],
\label{eq:action}
\end{equation}
where  $D_a=\nabla_a+iqA_a$, and
$F_{ab}=\nabla_aA_b-\nabla_bA_a$.

We impose spherical symmetry and foliate the spacetime by outgoing
affine-null hypersurfaces,
\begin{equation}
ds^2=-V(u,\lambda)\,du^2-2\,du\,d\lambda
+r^2(u,\lambda)\,d\Omega^2 
\label{eq:metric}
\end{equation}
where $d\Omega^2$ is the line element of the unit sphere metric.
Each surface $u=\mathrm{const}$ is an outgoing null cone with a regular
vertex at $\lambda=0$; $u$ is central proper time and $\lambda$ is an affine
parameter along the generators of the null cones.
We
choose the null gauge $A_a dx^a=\alpha(u,\lambda)\,du$ and fix its residual
freedom by $\alpha(u,0)=0$.  Regularity at the center requires
$r=\lambda+O(\lambda^3)$ and $V=1+O(\lambda^2)$.
The asymptotic redshift factor
$
H(u)=\lim_{\lambda\to\infty}r_{,\lambda}(u,\lambda),
$ 
relates central proper time to Bondi time $u_B$ at $\mathfrak{I}^+$ via $u_{B,u} = H^{-1}$.  In collapsing evolutions,
$H\to0$ at a finite central time $u_E$, while $u_B$ grows without bound.
The null hypersurface $u=u_E$ is the event horizon \cite{Crespo:2019mcv},
and the numerical evolution approaches it from the exterior. 
Affine-null coordinates penetrate the event horizon  and allow the tracking of an apparent horizon, $\mathscr H_A$, by following the curve $r_{,\lambda}(u,\lambda_{\mathrm{AH}}(u))=0$ where $\lambda_{\mathrm{AH}}(u)$ is its locus 
\cite{Madler:2018bmu}.

On each null hypersurface, the enclosed charge and Misner--Sharp mass
\cite{Misner:1964je} are
$Q=r^2\alpha_{,\lambda}$ and
$M=r(1-g^{ab}\nabla_a r\nabla_b r)/2$.  
In the infinite limit,  $\lambda\rightarrow \infty$, meaning  towards future null infinity $\mathfrak I^+$, the two define 
the Bondi mass $M_B(u)$ and {\it electromagnetic scalar Bondi charge} $Q_B(u)$.
In spherical symmetry, the Maxwell field carries no radiative degrees of
freedom; radiation at $\mathfrak I^+$ is characterized by the scalar field monopole
$C=\lim_{r\to\infty}r\Phi$ and the news
$N=dC/du_B$. 
As the electromagnetic gauge is fixed at the
center, $\alpha_\infty=\lim_{r\to\infty}\alpha$ does not vanish in
general \cite{Madler:2025ibn}. 
The physical fluxes at $\mathfrak I^+$ involve the 
gauge-covariant news function,
$\mathcal N=N+iqH\alpha_\infty C$, and
the corresponding balance
laws of $M_B$ and $Q_B$ in a  Bondi frame are  \cite{Madler:2025ibn}
\begin{subequations}\label{eq:dMdu_dQdu}
\begin{align}
\frac{dM_B}{du_B}
&=-4\pi|\mathcal N|^2,\label{eq:dMBduB}\\
\frac{dQ_B}{du_B}
&=-4\pi q\,\operatorname{Im}
\left(\overline C\,\mathcal N\right).\label{eq:dQduB}
\end{align}
\label{eq:balance-laws}
\end{subequations} 
An additional independent balance law  at $\mathfrak{I}^+$ is 
\begin{equation}\label{eq:cNP}
    \frac{d C_{\rm NP}}{du} = -iq\left(\alpha_\infty C_{\rm NP} -\frac{1}{2H}CQ_B\right)\,,
\end{equation}
where  $C_{\rm NP}(u_B) = -\lim_{r\rightarrow\infty}[r^2(r\Phi)_{,r}]$  is the  Newman--Penrose (NP) charge of the charged scalar field and which is conserved iff  $q=0$ 
\cite{Newman:1968uj,Gomez:1994ns}.

A non-vanishing field $\alpha_\infty$ is the electromagnetic analogue of the redshift $H$.
From a general frame $H\neq0$ at $\mathfrak I^+$, the unique Bondi frame with $H=1$ is obtained by the   coordinate transformation  
between $u$ and $u_B$.
In a similar way,  the {\it electromagnetic Bondi gauge} 
given by  $\alpha=0$  is obtained with the   gauge transformation $\Lambda(u_B) = \int_0^{u_B}H\alpha_\infty du^\prime_B$ at $\mathfrak I^+$, which is a result of the remaining gauge freedom of the charged scalar field. 
The variables $\mathcal O = \{\mathcal N, C, C_{\rm NP}\}$ are transformed like ${\mathcal{O}|_{\alpha\neq0}\rightarrow \mathcal{O}_0:=e^{iq\Lambda}\mathcal{O}|_{\alpha\neq0}}$ from a gauge $\alpha \neq0$ to $\alpha=0$ gauge. 

\textit{Numerical implementation and initial data.---}
The field equations resulting from \eqref{eq:action} and \eqref{eq:metric} can be written as a  hierarchy of radial equations
supplemented by an evolution equation for $\Phi$
\cite{Madler:2025ibn}.  
The numerical method follows the code architecture as presented in \cite{Madler:2024kks}, where we introduced a grid function $\lambda(x)$ with $x\in[-1,1]$ such that $\lambda(-1)=0$ is at the vertices of the null cones and  $\lim_{x\rightarrow1}\lambda(x)=\infty$ corresponds to future null infinity, $\mathfrak{I}^+$. 
At each value of $u$, we solve the radial hierarchy from the regular center
at $x=-1$ to $\mathfrak I^+$ at $x=1$ using Chebyshev--Gauss--Lobatto
collocation.  
The scalar-field evolution is integrated in $u$ using a fourth-order
Runge--Kutta method; a third-order scheme provides an independent convergence
check. The formally singular terms at the vertex are
evaluated with ninth-order Taylor expansions.
Further details and convergence analysis are given in Sec.~\ref{app:formalism}.

The freely specifiable initial datum is the complex scalar profile for $\hat\Phi$ on the
outgoing null cone $u=0$. 
The initial data are chosen horizonless so that the scalar field extends to $\mathfrak{I}^ +$ and in particular $H\neq0$ for $u=0$.
We investigate initial data of the type 
$\Phi(0,x) = f(x)e^{i\omega x}$, where the  real function $f(x)$ constitutes a
 Gaussian pulse centered at $x=0$,  with an amplitude $\mathcal{A}$ and {width  $\sigma_G=0.25$}. 
We test the evolution by increasing the spectral resolution and monitoring
the supplementary field equations, charge conservation, and the
asymptotic mass- and charge-balance laws. 

\begin{figure}[ht!]
    \centering
\includegraphics[height=0.18\textheight]{ 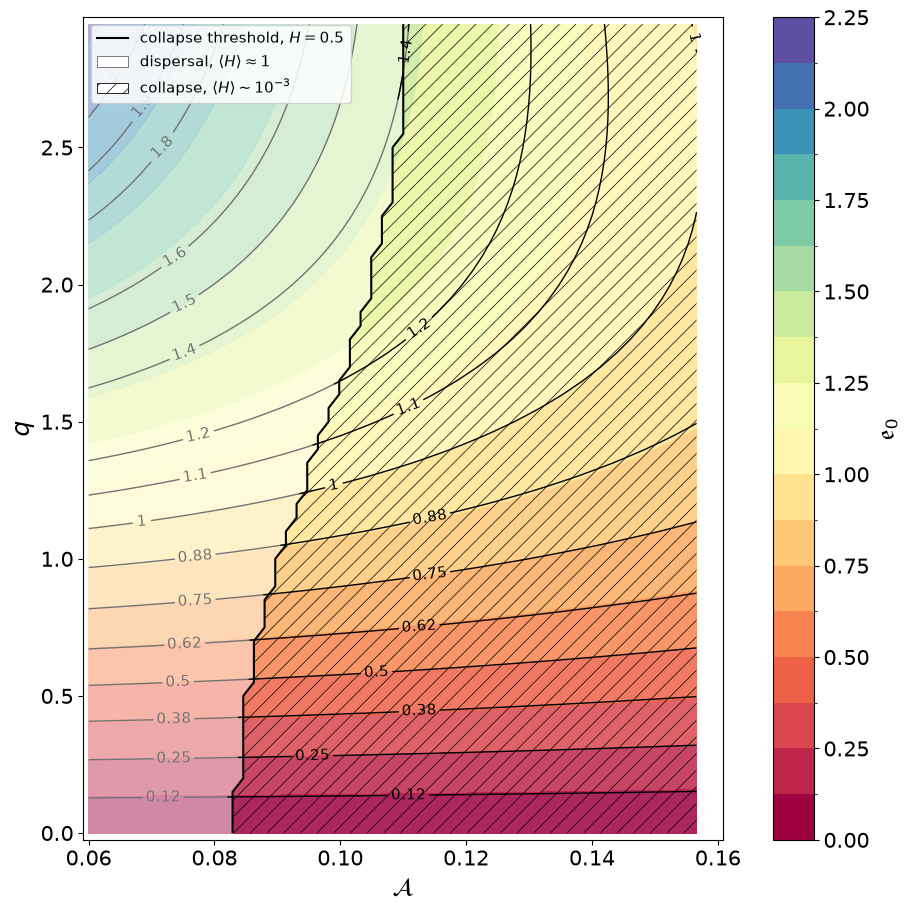}
\includegraphics[height=0.18\textheight]{ 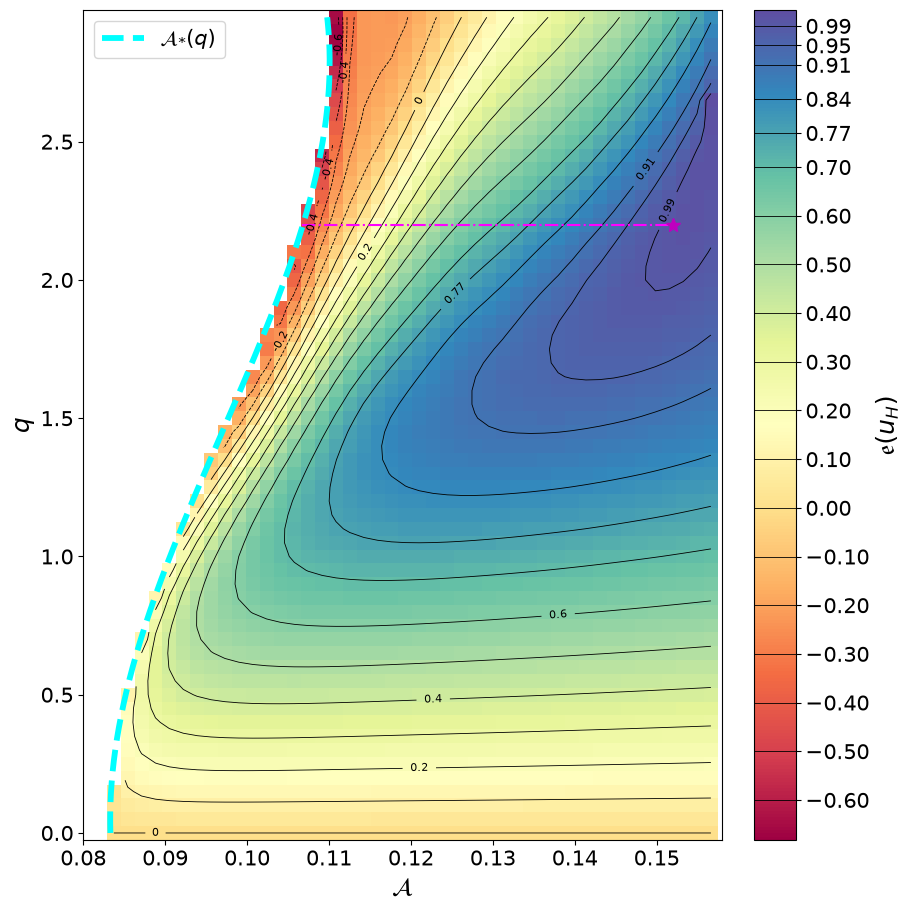}
        \caption{Charge-to-mass ratio $\mathfrak{e}$ over the $(\mathcal{A},q)$ plane: initial $\mathfrak{e}$ overlaid with the final redshift $H$ (left) and final $\mathfrak{e}$ in the black-hole region (right). Cyan dashed line: polynomial fit of the threshold $\mathcal{A}_*(q)$. Magenta dashed-dotted line, $q=2.2$ scan. Fields at $\mathfrak{I}^+$ for the magenta star initial data are in Fig.~\ref{fig:ringing}.
    }\label{fig:landscape_ini_fin}
\end{figure}

\textit{Parameter spaces of initial and final data.---}
We have  made extended scans in the parameter spaces for the phase $\omega$, the amplitude $A$, and non-negative gauge charge $q$.
In those scans, the two parameters $q$ and $\mathcal{A}$ were equidistantly varied, while the others are kept constant. 
 The left panel of Fig.~\ref{fig:landscape_ini_fin} shows in 
 color the distribution of the initial charge-to-mass ratio $\mathfrak{e}\equiv Q_B/M_B$ as measured at null infinity and as a function of Gaussian amplitude and gauge coupling charge. 
It is seen  that the initial values of $\mathfrak{e}$ are non-negative and that a significant amount of initial data are superextremal with $\mathfrak{e}>1$. 
Their corresponding Gaussian profiles  
were evolved for each pair $(q, \mathcal{A})$ and the outcome of the evolution monitored at $x=1$ was found to be  either: (i)~dispersal to flat space with $H\rightarrow 1$ for $u\rightarrow \infty$ or (ii)~formation of an  event horizon  with  $H\rightarrow 0$ for {$u\rightarrow u_H>0$} 
which stops the code. 
We denote by $u=u_H$ the last null cone on which \eqref{eq:dMBduB} holds before the run breaks down.
The final redshift distribution $H$  is overlaid in the left plot of Fig.~\ref{fig:landscape_ini_fin}, where in the left unshaded region $H\approx 1$ and in the right shaded region $H\sim10^{-3}$.
The threshold curve $\mathcal{A}_*(q)$, where $\mathcal{A}_*$ is the threshold amplitude, separates the two regions  determined by a bisection search in $\mathcal{A}$ while keeping $q$ constant.
The right panel of Fig.~\ref{fig:landscape_ini_fin} displays the final distribution  of $\mathfrak{e}$ 
in the black hole region, where $H\approx 10^{-3}$. 
We observe that $\mathfrak e$ varies between negative values (red)  and  positive values (blue) approaching unity. 
Values with $|\mathfrak{e}|\approx 0$  shown by a yellow color in the right panel of Fig.~\ref{fig:landscape_ini_fin}. 
This yellow region appears at the line $q\approx0$ and can furthermore be found along the black hole threshold up to $q\sim 1.2$ as well as  along the curve connecting (approximately) the  pairs $(\mathcal{A}, q) = (0.097, 1.4)$ and $(\mathcal{A}, q) = (0.13, 2.9)$.
Between the threshold and this  curve  we find negative values for the charge-to-mass ratio, whose values can go down to $\mathfrak{e}\approx-0.6$ near the threshold. Charge reversal has also been observed in earlier simulations of charged-scalar collapse \cite{Torres:2014fga}.
Finally, we find a region in the upper-right around $(\mathcal{A}, q) \approx (0.155, 2.25)$, bound to the right by initial data containing horizons, in which black holes approaching extremality are found.
    
\textit{Critical scaling and apparent horizon diagnostics.---}

Near the threshold, we expect $\mathfrak e\to0$ for the critical collapse of charged scalar fields \cite{Gundlach:1996vv,Hod:1996ar}, and   for 
$0<q<1$, our measurements at $\mathfrak I^+$ recover the expected scaling exponents $\gamma=0.37$ for the Bondi mass and $\delta=0.88$  for the Bondi charge. Nevertheless, the nonzero near-threshold ratios at larger $q$ in Fig.~\ref{fig:landscape_ini_fin} suggest that the evolutions may not have reached stationarity, raising doubts about the inferred black-hole charge-to-mass ratios even at larger amplitudes $\mathcal A$.

\begin{figure}[ht!]
    \centering
    \includegraphics[width=0.32\textwidth]{ 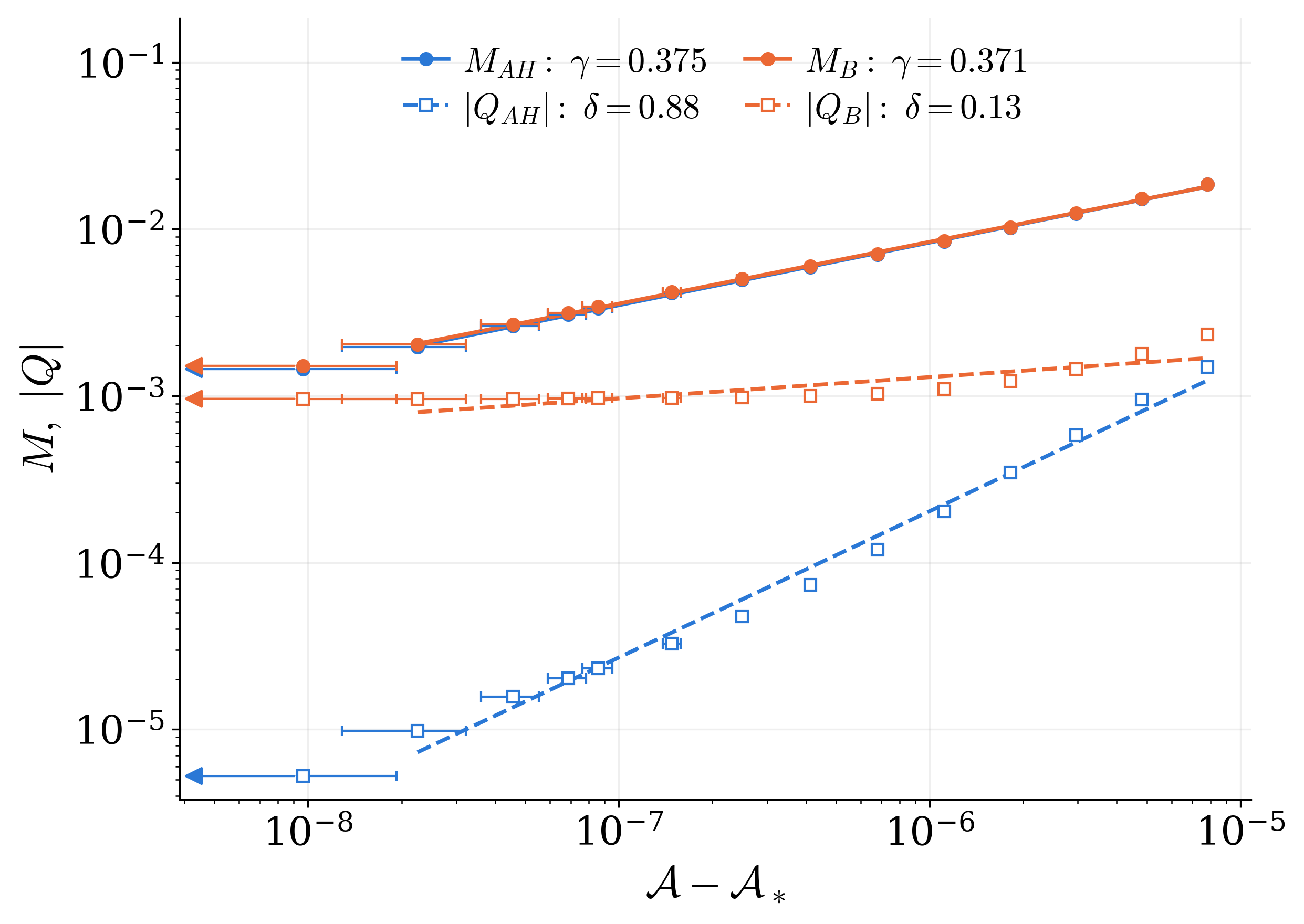}   
    \includegraphics[width=0.32\textwidth]{ 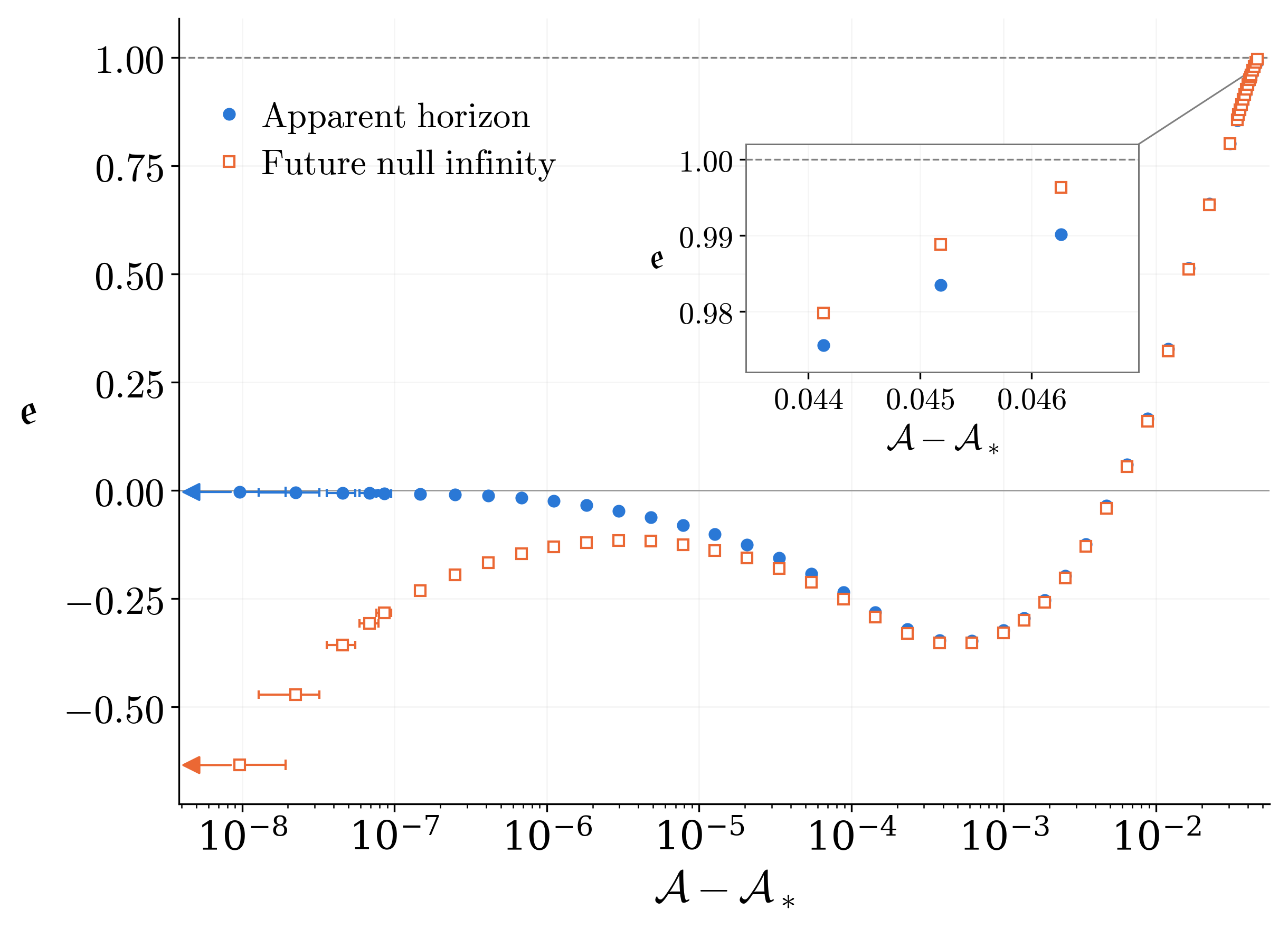} 
    \caption{Scaling laws and asymptotic behaviors for $q=2.2$ at $\mathscr H_A$ (blue) and  $\mathfrak{I}^+$ (orange). Upper: mass (filled circles) and charge (open squares) with their fitted slopes indicated. 
    Lower: Charge-to-mass ratio vs. $\mathcal{A}-\mathcal{A}_*$ show their agreement near extremality and a disagreement at the black hole threshold, where $\mathfrak e$ vanishes at $\mathscr H_A$ and is nonzero at $\mathfrak{I}^+$.}\label{fig:rAH_scri_scaling}
\end{figure}

To resolve this, we treat the apparent horizon  as in \cite{Madler:2024kks}: it is located on the first null cone not reaching $\mathfrak I^+$, where $r_{,\lambda}(\lambda_{\rm AH})=0$ for $\lambda_{\rm AH}>0$.
There, the enclosed charge $Q_{\rm AH}$ and area distance $r_{\rm AH}$ give the Poisson--Israel mass $M_{\rm AH}=(r_{\rm AH}^2+Q_{\rm AH}^2)/(2r_{\rm AH})$ \cite{Poisson:1989zz}.
We compare the apparent horizon variables with the Bondi variables, $M_B$ and $Q_B$  on the last physically reliable null cone $u_H$ extending to $\mathfrak I^+$.
A fine scan along  $q=2.2$ (Fig.~\ref{fig:landscape_ini_fin}), shows in Fig. \ref{fig:rAH_scri_scaling} (upper panel) that
$\gamma_H=0.375$ and $\delta_H=0.88$ over the resolved scaling range.
The Bondi mass closely follows the apparent horizon mass, with $\gamma_B=0.371$, whereas the last recorded Bondi charge exhibits an effective slope of only $0.13$.

This discrepancy is consistent with continued charge transport in the exterior scalar field when the asymptotic evolution terminates. 
The balance laws \eqref{eq:dMdu_dQdu} explain why accurate mass scaling does not guarantee accurate charge scaling: mass loss is quadratic in the gauge-covariant news, whereas charge loss is bilinear in the monopole and news. Moreover, the critical charge becomes parametrically smaller than the mass. Faithful charge extraction therefore requires the remaining charge correction to be small compared with $|Q_{AH}|\propto|\mathcal A-\mathcal A_*|^{0.88}$, a condition increasingly stringent near threshold. With $M_B\simeq M_{AH}$, the ratio bias is approximately $(Q_B-Q_{AH})/M_{AH}$, amplifying even a small charge mismatch as the mass vanishes. A small redshift alone consequently does not establish that the charge has settled.

For larger values of $\mathcal A-\mathcal A_*$, the  charge-to-mass ratios at $\mathscr H_A$ and $\mathfrak I^ +$ agree closely 
(Fig. \ref{fig:rAH_scri_scaling}
). Crucially, this agreement extends into the region where both ratios approach unity, as resolved in the inset. The comparison directly validates the interpretation of the near-extremal configurations identified at $\mathfrak I^+$: their large asymptotic charge-to-mass ratios are accompanied by independently measured near-extremal horizon parameters.  
Thus,  the Bondi quantities can contain contributions from the exterior scalar field and need not characterize the black hole alone.

For a settled extremal RN black hole, the Bondi mass and charge coincide. 
The agreement of $\mathfrak e$ at $\mathfrak{I}^+$ and the $\mathscr H_A$ in Fig. \ref{fig:rAH_scri_scaling}  supports using measurements at $\mathfrak I^+$ to guide a systematic search for near-extremal solutions and the extremal limit. The difficulty encountered near the collapse threshold concerns resolving a charge that vanishes faster than the mass; it does not invalidate the asymptotic identification of near-extremal black holes in the region where the horizon comparison succeeds. 

\begin{figure}[ht!]
    \centering
    \includegraphics[width=0.5\textwidth]{ 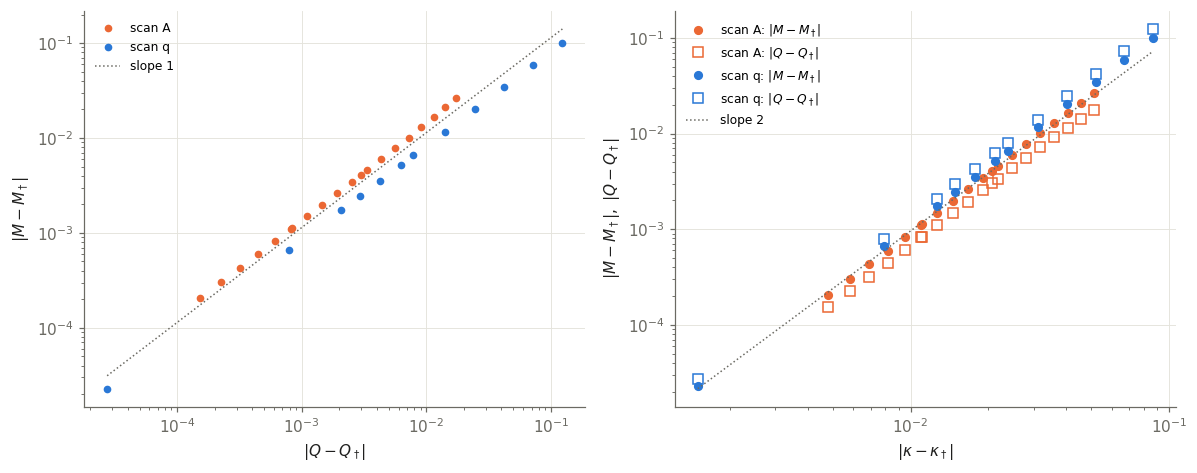}
    \caption{
    Bondi mass and charge scaling around the near-extremal reference solution (\(\dagger\)) over the resolved range. Dotted lines indicate slopes 1 (left) and 2 (right).}
   \label{fig:BHscaling}
\end{figure}

We examine the region of parameter space in which the evolution approaches closest to extremality, namely $q=2.2$ as shown in the right panel of Fig.~\ref{fig:landscape_ini_fin}. In particular, we varied separately either $\mathcal{A}$ or $q$ in the approach to $\mathfrak{e}\rightarrow 1$.
As discussed in~\cite{Angelopoulos:2026bez,Murata:2013daa}, one expects the mass and charge to scale as the second power of the black hole surface gravity, $\kappa \equiv (M^2-Q^2)^{1/2}r_+^{-2}$ with $r_+ \equiv M+\sqrt{M^2-Q^2}$.
In particular, we denote $\left\{M_\dagger,Q_\dagger\right\}$ as the mass and charge of our evolution 
closest to extremality, and then we expect nearby solutions to scale as
$|M-M_\dagger| \sim |\kappa-\kappa_\dagger|^2$ and $|Q-Q_\dagger| \sim |\kappa-\kappa_\dagger|^2$.
In Fig.~\ref{fig:BHscaling}, we display data for such solutions that indeed suggest this same near-extremal scaling.

\begin{figure}[ht!]
    \centering
    \includegraphics[width=0.4\textwidth]{ 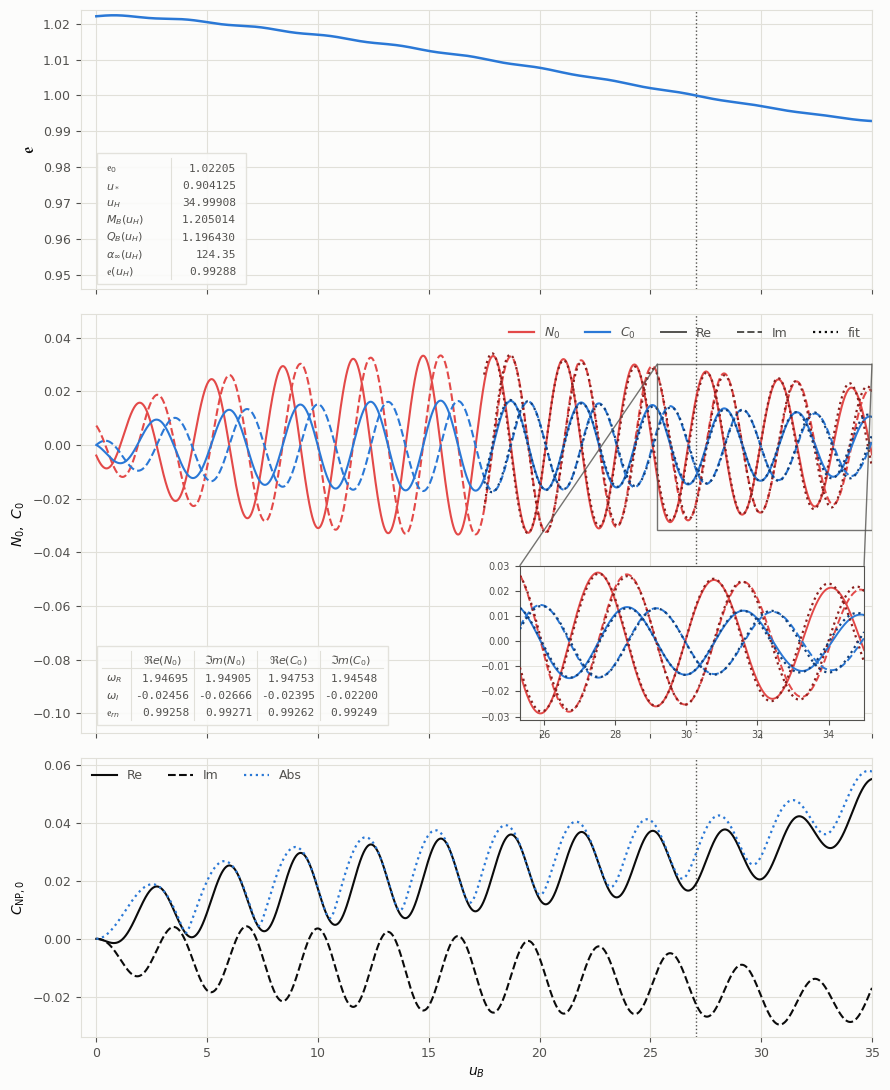}
    \caption{
    Evolution of $\mathfrak{e}$ (upper), news and monopole (middle), and NP charge (lower) versus Bondi time $u_B$ for initially superextremal ($\mathfrak{e}_0=1.022$), horizon-free data $(\mathcal{A},q,\omega)=(0.152,2.2,3)$, up to $u_H=35.0$. The dotted vertical line marks crossing of $\mathfrak{e}=1$. Upper table: $u_*$, $M_B$, $Q_B$, $\alpha_\infty$ and $\mathfrak{e}$ on the last null cone reaching $\mathfrak I^+$. Inset (middle): fit of a damped oscillation (dotted) to the tail.}
    \label{fig:ringing}
\end{figure}

\textit{Analysis at null infinity.---}
For near extremal final states, $\mathfrak{e}\approx1$,  in the black hole region, we were able to identify  the ringing of  zero-damped quasinormal modes.
    Fig.~\ref{fig:ringing} shows for the magenta-starred initial data in Fig.~\ref{fig:landscape_ini_fin} the $u_B$ evolution of $\mathfrak e$,  the news function, the scalar monopole and the NP charge evolution, in particular the latter is not constant.

Between $u_B\approx18$ and $u_H$,  where the news and monopole are decaying, we  fit a decaying oscillating function with 
a decay rate $\omega_I$ and angular frequency $\omega_R$. 
The corresponding fit values are shown in an inset table of the respective plots.
For $|qQ_B|>1/2$, the monopole of a charged scalar field on a near-extremal RN black hole rings in zero-damped quasinormal modes with $\omega_R \to qQ_{\rm RN}/r_+$ and $\omega_I \simeq -\kappa_H/2$ for the fundamental mode in the electromagnetic Bondi gauge \cite{Richartz:2014jla,Zimmerman:2016qtn}. From the angular frequency of the zero damped mode in the news function and $q$, we determine  charge-to-mass ratios of the black hole via $\mathfrak{e}_{\rm RN}   =2q\omega_R(\omega^2_R+q^2)^{-1}$ with their resulting values given in the tables of the plots. 
The theoretical values  of the quasinormal modes  of a charged scalar field with  gauge charge $q=2.2$ of a near extremal black hole with mass  $M_B(u_H)$ and a charge-to-mass ratio of $0.993$ are 
$\omega_R=1.96$ and $\omega_I=-0.04	$
\cite{Richartz:2014jla,Richartz:2015saa}. 
The excellent agreement between the measured and the theoretical value of $\omega_R$, provides further evidence that the evolution to near extremality is physical.
The difference of $\sim35\%$ for $\omega_I$, we attribute to the fact that the black hole is still dynamical and unsettled.

Near an event horizon the redshift decays like $e^{-\kappa_H u_B}$, \cite{Hawking:1975vcx,Crespo:2019mcv}, from which we determine the surface gravity, $\kappa_H$, of the forming black hole.  
From the exponential decay of the redshift and the extracted final value of the Bondi mass, we can determine the charge-to-mass ratio $\mathfrak{e}(H, M_B)$ from the root of the RN surface gravity formula $\kappa(\mathfrak{e}, M_B) = sM_B^{-1}(1
+s)^{-2}$ with $s=\sqrt{1-\mathfrak{e}^2(H, M_B)}$. 
The resulting surface gravity is $\kappa_H(u_H) = 0.079$ and  $\mathfrak{e}( H, M_B)=0.9928$     compares excellently with the extracted values of $\mathfrak e$ from the quasi-normal modes.

\textit{Discussion and Outlook.---}
In a coordinate chart of a {\it four dimensional spacetime} covering the vertices of a family of  null cones extending to  null infinity, we successfully identified horizon free initial data  that evolve up until the formation of an event horizon of near extremal black hole with $\mathfrak{e} =0.993$. 
This was achieved by extensively scanning the space of Gaussian initial data and a posterior analysis  at  $\mathfrak I^+$. 
This simulate-first and agnostic, ``astronomy-type"  data analysis  at $\mathfrak I^+$  is in spirit different to the approaches in \mbox{\cite{Gelles:2026bix,Mittal:2026ryg}} where near-extremal black holes are generated `inside-out' with null hypersurface evolutions not covering null infinity. 
In particular, \cite{Gelles:2026bix} forms extremal black holes from data not covering the vertex and starts out with an initially superextremal RN spacetime that is subject to scalar field perturbations. 
The authors of   \cite{Mittal:2026ryg} evolve initial data having a regular vertex but a finite size null cone not covering $\mathfrak I^+$ in a five dimensional spacetime.

Our scans revealed a region of slightly superextremal initial data that promptly collapse into near-extremal black holes. Since $Q_B$ is conserved for a neutral field, $\mathfrak e$ at $\mathfrak I^+$ can only grow in the model of \cite{Angelopoulos:2026bez}; here charge radiation \eqref{eq:dQduB} lets it decrease through unity, so extremality is approached from above, as in \cite{Gadioux:2025unn,Gelles:2026bix,Mittal:2026ryg} and unlike \cite{East:2025nfb}.
The black-hole threshold, located via the redshift, is well fitted by a polynomial relation $\mathcal A_*(q)$, along which we extract the Choptuik exponent $\gamma=0.37$ for the Bondi mass and, for $q<1$, the charge exponent $\delta=0.88$.
For $q>1$, the charge-to-mass ratio on $u=u_H$ instead stays finite near threshold and can be negative (Fig.~\ref{fig:landscape_ini_fin}), as allowed by \eqref{eq:dQduB}, whereas at the apparent horizon $Q\to0$ with $\delta=0.88$ (Fig.~\ref{fig:rAH_scri_scaling}).
We attribute this to a Bondi charge that has not settled when the run breaks down: focusing of the null rays near horizon formation stiffens the equations before the exterior scalar field has carried off its charge.

The radiation fields at $\mathfrak I^+$ predict the event-horizon parameters before the horizon forms: our analysis of the redshift decay and the quasinormal frequencies in the news and monopole both accurately indicate a near-extremal horizon  formation.
Since our runs end abruptly, other defining properties  of extremality remain out of reach.
Close to an extremal RN horizon, the spacetime is locally  a Robinson-Bertotti spacetime with an AdS$_2\times S_2$ geometry.
This  throat leaves imprints in the zero-damped quasinormal modes at $\mathfrak I^+$ \cite{Zimmerman:2016qtn}; our runs are too short to separate this damping from numerical artefacts of horizon formation.
Likewise, for a neutral field on exact extremal RN, the Aretakis horizon charge equals the NP charge of the Couch--Torrence dual field at $\mathfrak I^+$ \cite{Angelopoulos:2018yvt,Cvetic:2018gss}.
As we have shown, $C_{\rm NP}$ is not conserved for a charged field, its relation to a horizon quantity is an open question.
In Fig.~\ref{fig:ringing}, $|C_{\rm NP}|$ grows faster after $\mathfrak e$ crosses unity, consistent with a transient Aretakis-type instability 
for $v\lesssim2\pi/\kappa_H\approx80$ \cite{Angelopoulos:2026bez}, although numerical artefacts cannot be excluded.
We note that our final values of $\mathfrak{e}$ and $M_B$ are consistent with the recent bound $qr_+=qM(1+\sqrt{1-\mathfrak{e}^2})>1/3$  for gluing  extremal RN spacetimes \cite{Schneider:2026zta}.
In $3+1$ dimensions, our results are the first to establish a dynamical route from regular initial data towards near-extremal black holes with $\mathfrak e\simeq0.993$, which are independently identified through redshift decay and quasinormal ringing at future null infinity. 
The emerging horizon geometry is thus encoded in asymptotic observables, connecting fully nonlinear collapse to black-hole spectroscopy. This makes $\mathfrak I^+$ a numerical laboratory for the extremal limit and brings the third law of black-hole mechanics within reach of direct tests through the radiation produced in collapse.\\

 \textit{Acknowledgments.---} All authors gratefully acknowledge the hospitality of the Institute for Computational and Experimental Research in Mathematics (ICERM) at Brown University, where this work was initiated. E.G. also acknowledges financial support from SeCyT-UNC and CONICET. T.M. and R.G acknowledge support from the FONDECYT REGULAR with No. 1262002 and No. 1262349 of the Chilean national funding agency ANID.
 T.M. thanks to discussions and support from R. Arias, V. Barr\'ia, J. Frauendiener, C. Fischer, P. Jofré, C. Stevens,  and  J. Winicour, and  the hospitality of the University of Canterbury and University of Otago, where part of this work was carried out, is acknowledged. 
 This work was also supported by the National Science Foundation via grant PHY-2409407,
and used computational resources provided by
the Advanced Cyberinfrastructure Coordination Ecosystem: Services \& Support (ACCESS) program.
We also acknowledge the use of AI technology from OpenAI and Anthropic, which helped in the numerical data analysis and the generation of some of the graphics.

{\it Data availability.---} The data supporting  this work are not publicly available upon publication because it is not technically feasible and/or the cost of preparing, depositing, and hosting the data would be prohibitive within the terms of this research project. The data are available from the authors upon reasonable request.

\bibliography{refs}

\begin{appendix}
\section{Affine-null metric formulation for a charged massless scalar field} \label{app:formalism}
The numerical code is based on the characteristic formulation of the Einstein-Maxwell-scalar field equations in affine null coordinates presented in \cite{Madler:2025ibn} with the spherically symmetric line element
\begin{equation}
    ds^2 = \Big[-V(w, \lambda)dw + 2\epsilon d\lambda\Big]dw 
    +r^ 2(w,\lambda)q_{AB}dx^Adx^ B,
\end{equation}
where $\epsilon=-1$ represents a foliation with respect to a family of outgoing null cones $w=u=const$ and $\epsilon=+1$ a family of ingoing null cones $w=v=const$. The null cones have their vertices at the central worldline where $\lambda=0$ and the geometry is locally inertial so that
$V(w,0)-1=r(w,0)=0$.

To cast the main field equations of the Einstein-Maxwell-scalar field equations into the typical hierarchy of hypersurface-evolution equations use the definitions 
\begin{subequations}\label{eq:hyp_vars}
\begin{align}
    Z =& Vrr_{,\lambda}+2 \epsilon rr_{,w}-\lambda\,,\\\
    \mathcal{L} =& 2r(\Phi_{,w} + i q\alpha \Phi)+ r\epsilon V\Phi_{,\lambda}\,, \\
    Q=&-\epsilon r^2 \alpha_{,\lambda}.
\end{align}
\end{subequations}
The new fields \eqref{eq:hyp_vars} transform the field equations into a system of hypersurface-evolution equations similar to those  presented in \cite{Madler:2025ibn}.

For simulations covering null infinity, we introduce the grid function $\lambda = 2A(1+x)/(1-x)$, and for those with a finite size null cone $\lambda = A(1+x)$. 
A compactified grid function will lead to singular terms in the field equations \cite{Madler:2025oiy}, we regularize the equations using the fields
\begin{equation}\label{eq:def_RWhatPhi}
    R= y r\;,\;
    W= y V\;,\;\hat\Phi=\Phi/ y\;,\;\ell\equiv y\lambda\;,\;\nu\equiv\lambda_{,x} y^2,
\end{equation}
where the values $ y$, and $\ell$ depend on the choice of grid function, see Table~\ref{tab:grid_maps}.
With the definitions \eqref{eq:def_RWhatPhi} the hypersurface-evolution equation system of \cite{Madler:2025ibn} (eq.~3.5) is 
\begin{subequations}\label{hyp_ev_num}
\begin{align}
0 &= R_{,xx} + \frac{\kappa R}{2}\,\big| D\hat\Phi \big|^{2}\,, \\
Q_{,x} &= 4\pi q R^{2}\, \Im m\left( \overline{\hat\Phi}\,\hat\Phi_{,x} \right)\,, \\
\alpha_{,x} &= -\epsilon\nu\,\frac{Q}{R^{2}}\,, \\
Z_{,x} &= -\frac{\kappa\nu}{8\pi}\,\frac{Q^{2}}{R^{2}}\,, \\
\mathcal{L}_{,x} &= -\frac{\epsilon\,(\ell +  y Z)}{R}\,D\hat\Phi - i\,\frac{\epsilon\nu q Q \hat\Phi}{R}\,, \\
W_{, x x} &=-\frac{2 \nu}{R}\left(\frac{\ell+ y Z}{R}\right)_{, x}+\frac{\kappa \nu^2  y Q^2}{4 \pi R^4}-\frac{\kappa \epsilon \nu}{R} \Re e(\overline{D \hat{\Phi}} \mathcal{L})\,,\\
\hat\Phi_{,w} &= \frac{\mathcal{L}}{2R} - i q \alpha \hat\Phi - \frac{\epsilon W}{2\nu}\, D\hat\Phi\,,
\end{align}
\end{subequations}
where $D\hat \Phi =  y\hat\Phi_{,x}+ y_{,x}\hat \Phi.$
\begin{table}[h]
\centering
\begin{tabular}{l||c|c|c|c}
\hline\hline
grid & $\lambda(x)$ & $ y$ &   $\nu$ & $\ell$  \\
\hline
compactified & $\dfrac{2A(1+x)}{1-x}$ & $1-x$ &   $4A$ & $2A(1+x)$   \\
\hline
linear        & $A(1+x)$              & $1$   &    $A$  & $A(1+x)$     \\
\hline\hline
\end{tabular}
\caption{Grid functions for the two choices}
\label{tab:grid_maps}
\end{table}
Local validity of \eqref{hyp_ev_num} can be monitored with the linear combination of the supplementary and the trivial equation, 
$$E_{ww}+\frac{\epsilon}{2}VE_{w\lambda}=0,$$
of the Einstein equations giving
\begin{align}\label{eq:res_Eww}
0=& y R^2 Z_{,w}
 + \frac{\kappa\epsilon y}{4}\left[R^2|\mathcal{L}|^2
 - \frac{ y W Q^2}{4\pi}\right]
 +\frac{\epsilon}{2} W R^2
 \nonumber\\
 &
 - \frac{\epsilon}{2}\left[  y( y Z+\ell)^2
 + \frac{R^2}{\nu}( y Z+\ell)\left( y W_{,x} -  y_{,x} W\right)\right]\,,
 \end{align}
and the supplementary equation $E_w=0$ from the Maxwell equations
\begin{align}\label{eq:res_Ew}
    0 &= Q_{,w} 
    + \frac{\epsilon\, y W}{\nu}\,Q_{,x}
   + 4\pi q^2 R^2\alpha|\hat\Phi|^2\nonumber\\
   &
 + 2i\pi q R^2\Bigl(\hat\Phi\overline{\hat\Phi}_{,w} - \overline{\hat\Phi}\hat\Phi_{,w}\Bigr)\,.
\end{align}
The Misner-Sharp mass, $M=(r/2)(1-\nabla^a r\nabla_a r)$ \cite{Misner:1964je},  is calculated from the hypersurface equation
\begin{equation}
\begin{split}   
    M_{,x} =& \frac{\kappa R}{4\nu}\,(\ell+ y Z)\left| y\hat\Phi_{,x}+ y_{,x}\hat\Phi\right|^2\\
 &- \frac{Z_{,x}}{2\nu}\left( y R_{,x}- y_{,x}R\right)\,,
\end{split}
\end{equation}
with $M(w, -1) = 0$ 
and which determines the Poisson-Israel mass \cite{Poisson:1989zz,Poisson:1990eh} (aka reduced Hawking mass) via
\begin{equation}
    M_{PI} = M+\frac{ y Q^2}{2R},
    \label{eq:M_PI}
\end{equation}
we see that $M_{PI} = M$ at $x=1$.
The Ricci scalar $\mathcal{R}=g^{ab}R_{ab}$, can be calculated from
\begin{equation}\label{RS}
    \begin{split}
        r\mathcal{R} = \epsilon\kappa\,\Re e\big(\overline{\Phi}_{,\lambda}\mathcal{L}\big)
 = \frac{\epsilon\kappa\, y^2}{\nu}\,\Re e\Big[\big( y\overline{\hat\Phi}_{,x}+ y_{,x}\overline{\hat\Phi}\big)\mathcal{L}\Big]\,.
    \end{split}
\end{equation}
At null infinity, the redshift $H$ is given by ${H=R(w,1)/(4A)}$, the Bondi mass $M_B=M(w, 1)$, the Bondi charge $Q_B=Q(w,1)$, the asymptotic value of the electromagnetic potential $\alpha_\infty=\alpha(w,1)$, 
the radiative monopole $ {C} = R(w,1)\hat\Phi(w,1)$, 
the gauge covariant news function is found from 
$\mathcal{N} = (H/2)\mathcal{L}(w  ,1)$,
and  consequently the residual of the balance law of Bondi mass decay \eqref{eq:dMBduB} is given by
\begin{equation}\label{eq:res_MB}
\mathscr{R}({M_B}) :=
H\frac{dM_B}{dw}
-
\frac{\epsilon\kappa H^2}{8}|\mathcal{L}(w,1)|^2
.
\end{equation}
The coefficient $C_{\rm NP} = -\lim_{r\rightarrow\infty}[r^2(r\Phi)_{,r}]$  gives rise to the asymptotic balance equation 
\begin{equation}\label{eq:res_cNP}
\mathscr{R}(C_{\rm NP}) :=  \frac{d}{dw}C_{\rm NP} 
    +
    iq\left(\alpha_\infty C_{\rm NP}
    +\frac{\epsilon {C}Q_B}{2H}\right)
    =0\,,
\end{equation}
which corresponds in the uncharged case ($q=0$) to the conservation of the (complex) Newman-Penrose constant \cite{Newman:1968uj,Gomez:1994ns}. 

The boundary values of the field are fixed by regularity and gauge conditions at the origin. 
In particular,  the hypersurface fields $\alpha$, $R$, $Z$, $\mathcal{L}$, and $Q$ are trivial 
while $R_{,x} = 2A$, $W_{,x}=-1$, and $W=2$ at $x=-1$. Initial data on the initial null cone $w=0$ extending to $\mathfrak I^+$ is the Gaussian 
\begin{equation}\label{eq:ID_gauss}
\hat\Phi = \frac{\Phi}{ y} = \mathcal{A} e^{-\frac{(x-x_0)^2}{\sigma^2}}    e^{i\omega x}.
\end{equation}

The four residual equations \eqref{eq:res_Eww}, \eqref{eq:res_Ew}, \eqref{eq:res_MB} and \eqref{eq:res_cNP} allow us to define a global variable to validate the quality and error of the numerical implementation of the hypersurface evolution equations \eqref{hyp_ev_num}
\begin{equation}\label{eq:sum_res}\begin{split}
    \mathscr{R}_\Sigma =& \ell_2\left(E_{ww}+\frac{\epsilon}{2}VE_{w\lambda}\right)
    +\ell_2(E_w)\\
    &+|\mathscr{R}({M_B})|+|\mathscr{R}({Q_B})| + |\mathscr{R}({C_{\rm NP})}|\,,
\end{split}
\end{equation}
where $\ell_2 = \sqrt{\int_{-1}^ 1 |f(w,x)|^2 dx}$ is the spatial $\ell_2-$norm at $w=const$ and  $
\mathscr{R}({Q_B}):=E_w(w, 1)$ corresponds to the asymptotic balance law of the Bondi charge $Q_B$ at null infinity.

\section{Convergence behavior}
\begin{figure}[ht!]
    \centering
    \includegraphics[width=0.3\textwidth]{ 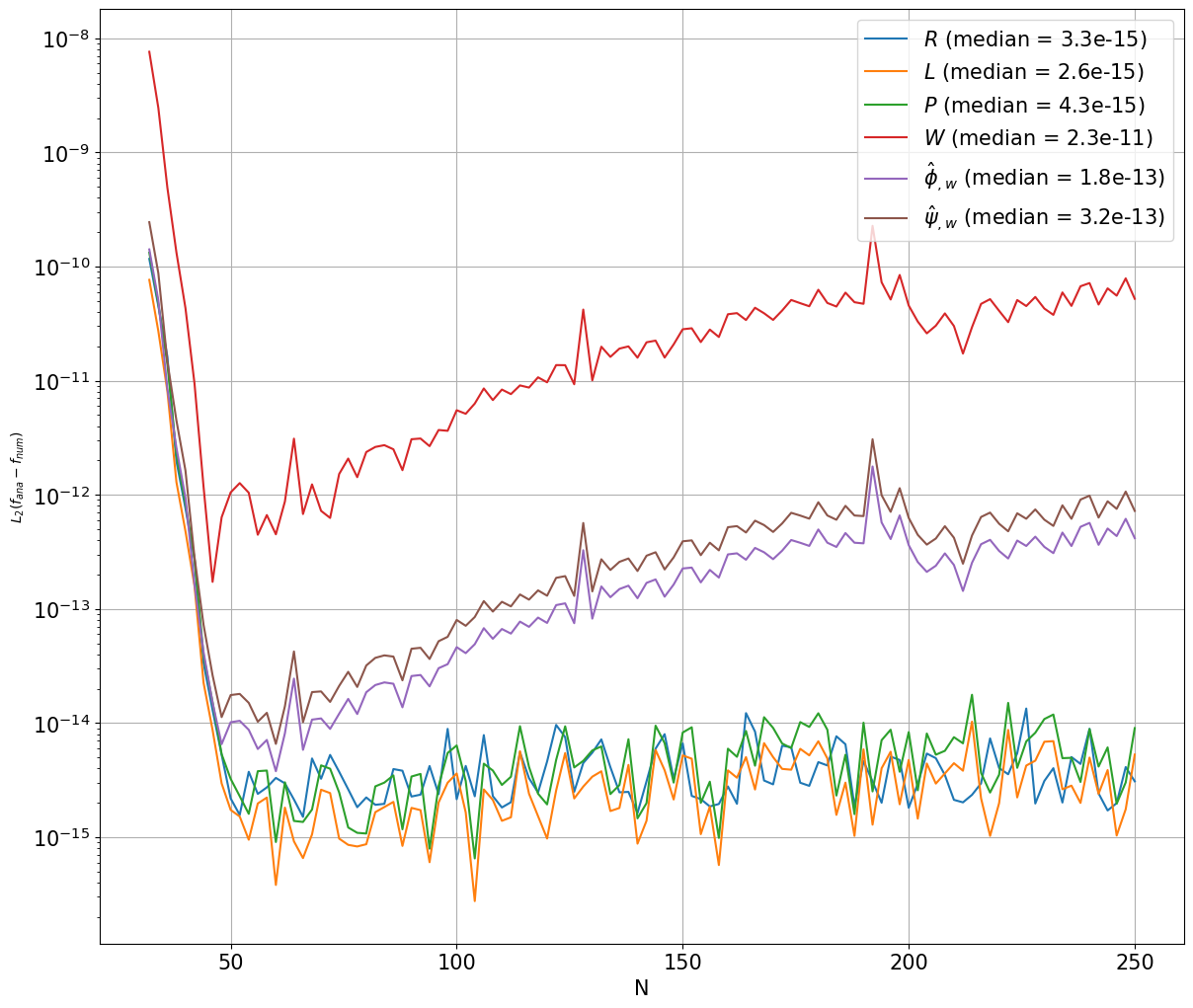}
    \caption{\textit{Convergence of hypersurface solver}.
       }\label{fig:conv_hyp}
\end{figure}

The numerical code and its result have been subject to robust numerical convergence studies. 
The solver of the hypersurface equations for the uncharged ($q=0$) case was tested against a complexified version of the analytic solution presented in \cite{Madler:2024kks}. Fig.~\ref{fig:conv_hyp} shows how the $\ell_2-$norm, $\ell_2(f-f_{ana})$, of the difference  of the hypersurface fields $f\in\{R, L, P, W , \hat \varphi_{,w}\, \hat \psi_{,w}\}$ in which $\mathcal{L} = L+iP$ and $\hat \Phi_{,w} = \varphi_{,w}+i \hat \psi_{,w}$ wrt its analytic solution, $f_{ana}$,  versus the number of spectral collocation points, $N$. It is seen that the field $W$, has for large $N$, the maximum error  saturating the round-off error  at $10^{-11}$.

\begin{figure}[ht!]
    \centering
    \includegraphics[width=0.4\textwidth]{ 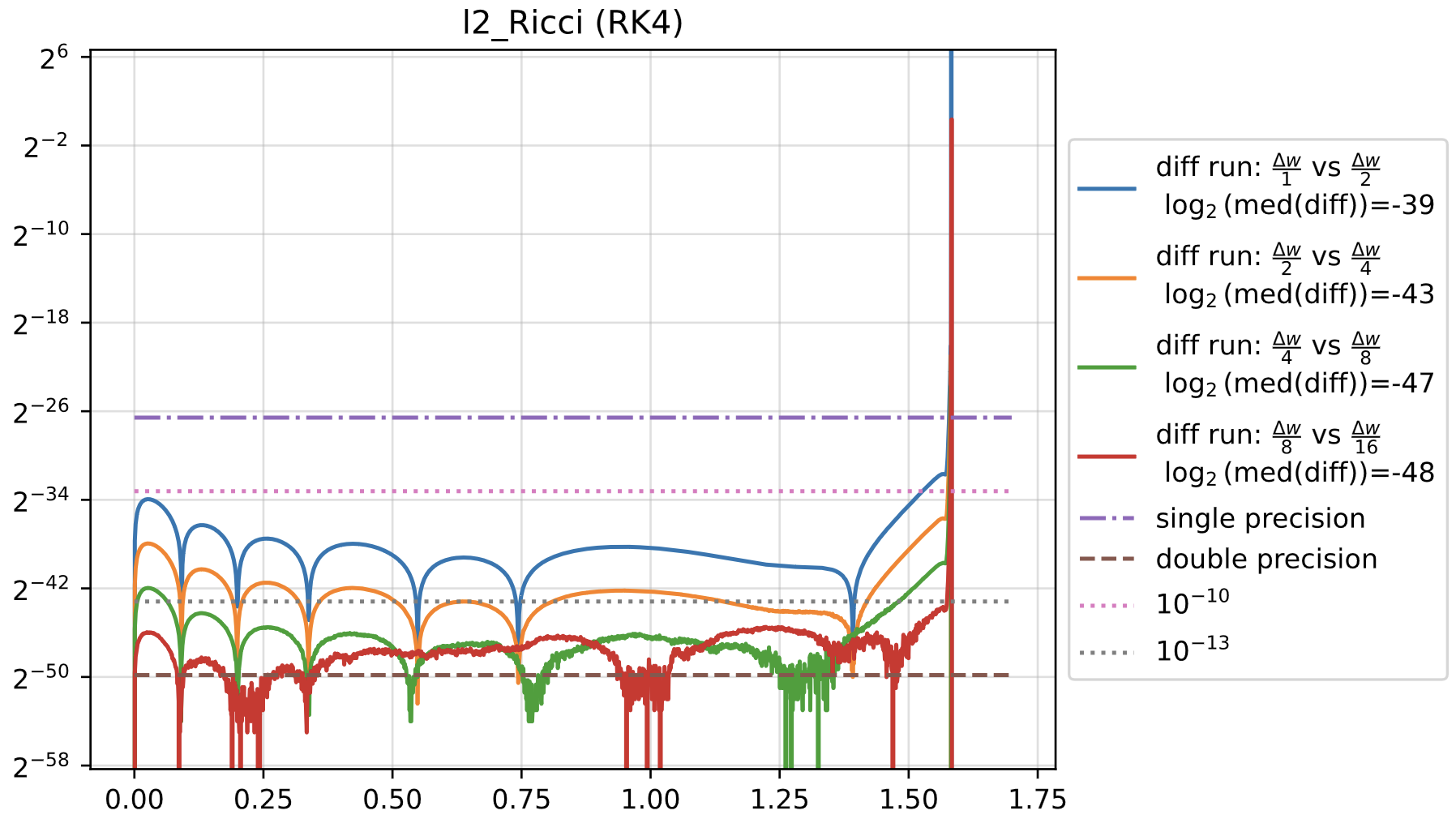}
    \caption{Convergence of the RK4 time integrator demonstrated at the spatial $\ell_2$-norm of the difference of two Ricci scalar profile at constant time $w$ but obtained from evolutions where the time steps differ by a factor $1/2$.
       }\label{fig:conv_time}
\end{figure}

The convergence behavior of the fourth-order Runge-Kutta  time integrator is illustrated at $u-$evolution of the $\ell_2-$ norm of the  Ricci scalar \eqref{RS} in Fig.~\ref{fig:conv_time} resulting from  5 runs where the time step was successively divided by two while keeping the spatial number of grid points constant. The consistency of the fourth order time integrator  is seen by comparing absolute errors between neighboring runs with $\Delta w/2^i$ and $\Delta w/2^{i+1}$, $i=0,1,2,...$, we can observe in Fig.~\ref{fig:conv_time} that the absolute error difference of  errors resulting from $(\Delta w, \Delta w/2) $, $(\Delta w/2, \Delta w/4) $ and $(\Delta w/4, \Delta w/8) $ differ by a factor $2^4=16$, while the comparison error of  $(\Delta w/8, \Delta w/16) $ is at the round-off floor $10^{-13}$. 

\begin{figure}[ht!]
    \centering
    \includegraphics[width=0.3\textwidth]{ 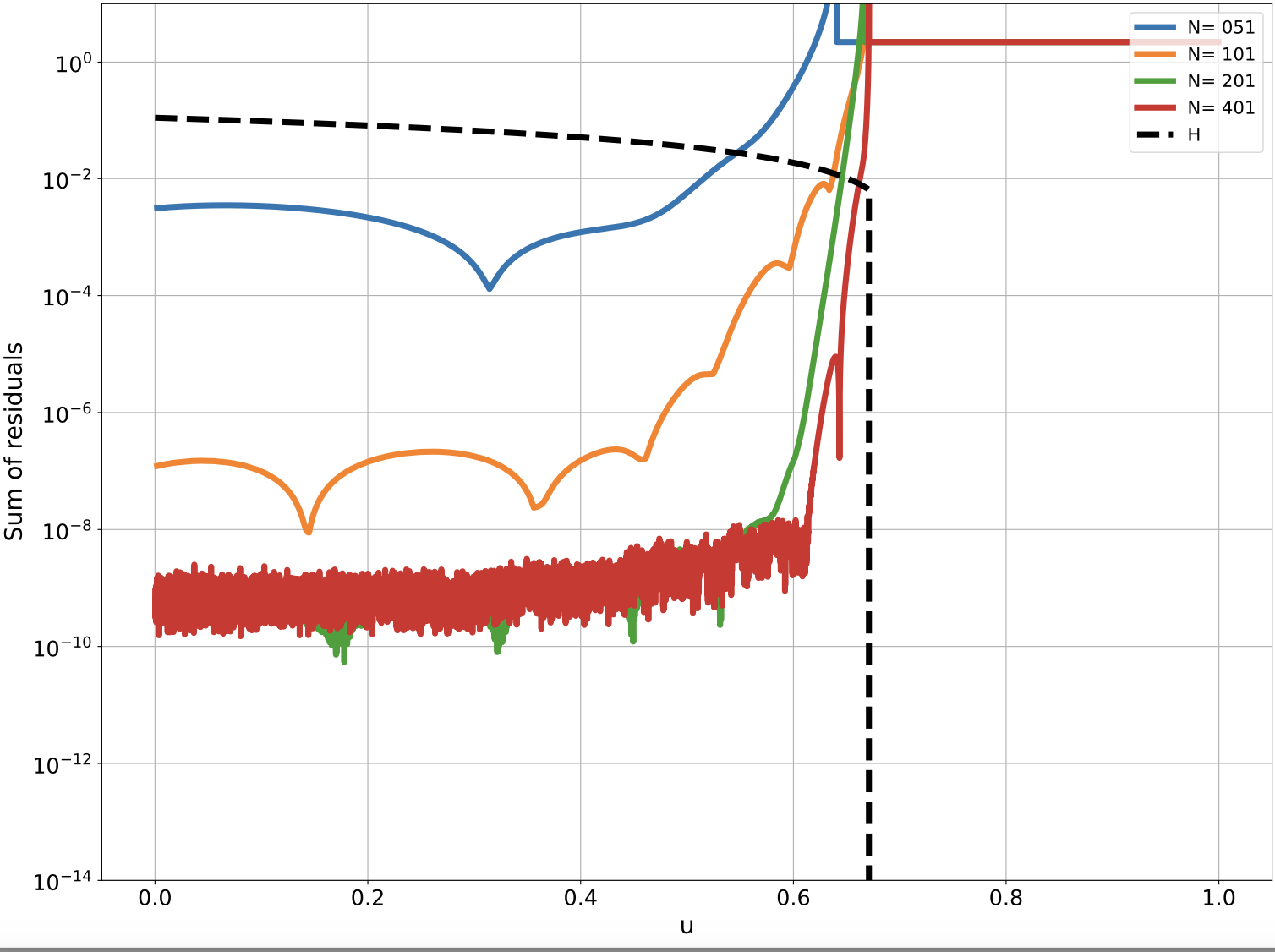}
    \caption{\textit{Convergence of residuals}.
       }\label{fig:conv_res}
\end{figure}

Physical consistency is monitored with  the temporal evolution of the sum of residuals \eqref{eq:sum_res}  is displayed in Fig.~\ref{fig:conv_res}, which shows exponential convergence to round-off saturation at $\sim 10^{-9}$ while doubling the resolution from 51 to 401 collocation points. At time $u\sim0.67$, the black hole forms, which is also seen at the dropping of the redshift $H$ while the residuals become infinite.

\end{appendix}
\end{document}